\documentclass[preprint,superscriptaddress]{revtex4}

\usepackage{color,psfig,graphics,psfrag,amsmath,amssymb,amsfonts,rotating}

\newcommand{\eps}{\varepsilon}

\begin{document}

\title{Risk Minimization through Portfolio Replication}

\author{Stefano Ciliberti} 

\affiliation{CNRS; Univ. Paris Sud, UMR8626, LPTMS, ORSAY CEDEX, F-91405
(France)}

\affiliation{Science \& Finance, Capital Fund Management, 6 bd Haussmann,
75009, Paris (France)}

\author{Marc M\'ezard} 

\affiliation{CNRS; Univ. Paris Sud, UMR8626, LPTMS, ORSAY CEDEX, F-91405
(France)}


\begin{abstract}
We use a replica approach to deal with portfolio optimization problems. A
given risk measure is minimized using empirical estimates of asset values
correlations. We study the phase transition which happens when the time
series is too short with respect to the size of the portfolio. We also study
the noise sensitivity of portfolio allocation when this transition is
approached.  We consider explicitely the cases where the absolute deviation
and the conditional value-at-risk are chosen as a risk measure.  We show how
the replica method can study a wide range of risk measures, and deal with
various types of time series correlations, including realistic ones with
volatility clustering.
\end{abstract}

\pacs{ }

\maketitle


\section{Introduction \label{sect:intro}}

The portfolio optimization problem dates back to the pioneering work of
Markowitz~\cite{marko} and is one of the main issues of risk management.
Given that the input data of any risk measure ultimately come from empirical
observations of the market, the problem is directly related to the presence
of noise in financial time series. In a more abstract (model-based)
approach, one uses Monte Carlo simulations to get ``in-sample'' evaluations
of the objective risk function. In both cases the issue is how to take
advantage of the time series of the returns on the assets in order to
properly estimate the risk associated with our portfolio. This eventually
results in the choice of the risk measure, and a long debate in the recent
years has drawn the attention on two important and distinct clues:  the
mathematical property of \emph{coherence}~\cite{artzner99}, and the noise
sensitivity of the optimal portfolio. The rational behind the first of these
issues lies in the need of a formal (axiomatic) translation of the basic
common principles of risk management, like the fact that portfolio
diversification should always lead to risk reduction. Moreover, requiring a
risk measure to be coherent implies the existence of a unique optimal
portfolio and a well-defined variational principle, of obvious relevance in
practical cases. The second issue is also a very delicate one. In a
realistic experimental set-up, the number $N$ of assets included in a
portfolio can be of order $10^2$ to $10^3$, while the length of a trustable
time series hardly goes beyond a few years, i.e. $T\sim 10^3$. A good
estimate of any extensive observable would require the condition $N/T\ll 1$
to hold, but this is rarely the case. Instead, the ratio of assets to data
points, $N/T$, will be considered as a finite number.

In this note we address analytically the risk minimization problem by
studying the dependence of the optimal portfolio on the ratio $N/T$ and on
other potential external parameters. We first assume that the real
distribution of returns is multinormal in order to keep the problem tactable
from the analytical point of view. Generalizations to more realistic returns
distributions are also presented. Our approch consists in writing down the
empirical estimate of the risk measure and then reformulating the problem
from the point of view of the statistical physics.  We work out the
analytical solution by means of the replica method~\cite{mepavi} and thus
get some insights on the optimal portfolios.  The analytical solution
confirms previous results on the existence of a phase
transition~\cite{kondor05}. The ratio $N/T$ plays the role of a control
parameter. When it increases, there exists a sharply defined threshold value
where the estimation error of the optimal portfolio diverges. A first
account of our method, limited to the Expected Shortfall risk measure, has
appeared in ref.~\onlinecite{es_replicas}. Here we give a more general
presentation, studying other risk measures and more realistic distributions
of returns. 

The paper is organized as follows. In section~\ref{sect:general} we
introduce the notations we will use throughout the paper and we formulate
the problem in its general mathematical form. In section~\ref{sect:AD} we
consider the case of the absolute deviation (AD) \cite{MAD}. We present the
replica calculation of the optimal portfolio and compute explicitely a noise
sensitivity measure introduced in ref.~\onlinecite{pafka02}.  In
section~\ref{sect:ES} we deal with portfolio optimization under Expected
Shortfall~\cite{artzner99, acerbi02}, which was shown to have a non-trivial
phase diagram~\cite{kondor05} and then studied
analytically~\cite{es_replicas}.  The striking point is that, for some
values of the external parameters of the problem, the minimization problem
is not well defined and thus cannot admit a finite solution. We investigate
here the same feature while considering realistic distribution of returns,
so as to take into account volatility clustering. The replica approch then
turns into a semi-analytic and extremely versatile technique.  We discuss
this point and then summarize our results in section~\ref{sect:conclu}.


\section{The general setting \label{sect:general}}

We denote our portfolio by ${\bf w}=\{w_1,\ldots w_N\}$, where $w_i$ is the
position on asset $i$. We do not impose any restriction to short selling:
$w_i$ is a real number. The global constraint induced by the total budget
reads $\sum_i w_i = N$, where, due to a later mathematical convenience, we
have chosen a slightly different normalization with respect to the previous
literature.  Calling $x_i$ the return of the asset $i$ and assuming the
existence of a well-defined probability density function (pdf) ${\sf p}
(x_1,\ldots x_N)$, one is interested in computing the pdf of the
\emph{loss} $\ell$ associated to a given portfolio, i.e.
\begin{equation}
  p_{\bf w}(\ell) = \int \prod_i d x_i \ {\sf p}(x_1,\ldots x_N) 
  \ \delta\left(\ell + \sum_{i=1}^N w_i x_i\right) \ .
  \label{eq:pell}
\end{equation}
The complete knowledge of this pdf would lead to the precise, though still
probabilistic, evaluation of the loss, thus allowing for a straightforward
optimization over the space of legal portfolios. This is actually a pretty
difficult task and one usually restricts to some characteristic of this pdf
(e.g. its first moments, its tail beahvior), so as to capture the
consequences of extremely bad events in the global loss.  The actual {\sf
p}$(x_1,\ldots x_N)$ is not known in general, and integrals like the one in
(\ref{eq:pell}) are usually estimated by time series, coming from market
oservations or synthetically produced by numerical simulations. Whatever the
chosen risk measure then, one typically faces cost functions (to be
optimized over all possible portfolios) like
\begin{equation}
  \text{risk}({\bf w};N,T,\lambda) = \frac 1T \sum_{\tau=1}^T   
  \mathcal{F}_\lambda\left[  \sum_{i=1}^N w_i x_i^{(\tau)}\right]
  \ ,
  \label{eq:risk}
\end{equation}
where $\{x_i^{(1)},x_i^{(2)},\ldots x_i^{(T)}\}$ is the whole time series of
the return $i$ and where we denoted by $\lambda$ other possible external
parameters of the risk measure. The best known example of risk measure is of
course the variance, as first suggested by Markowitz. In that case the risk
function is obtained by taking $\mathcal{F}_\lambda(z)=z^2$ in
(\ref{eq:risk}).
The evaluation of the variance implies an empirical evaluation of the
covariance matrix $\sigma_{ij}$ of the underlying stochastic process, and
the extremely noisy character of any estimation of $\sigma_{ij}$ has been
underlined a few years ago~\cite{laloux99, plerou99}.  However, recent
studies~\cite{pafka02,pafka03} have shown that the effect of the noise on
the actual portfolio risk is not as dramatic as one might have
expected. More in detail, a direct measure of this effect was introduced and
explicitely computed in the simplest case of $\sigma_{ij} = \delta_{ij}$.
In the next section, we compute the same quantity as far as the absolute
deviation of the loss is concerned.

In the statistical physics approach, one studies the limit $N,T\to\infty$,
while $N/T\equiv 1/t$ is finite. One introduces the partition function at
inverse temperature $\gamma$:
\begin{equation}
  Z^{(N)}_\gamma[t,\lambda;\{x_i^{(\tau)}\}]  = 
  \int \prod_{i=1}^N dw_i\  
  e^{-\gamma\;\textrm{risk} [{\bf w}; N, Nt,\lambda]}\ 
  \delta\left( \sum_{i=1}^N  w_i -N \right) \ ,
  \label{eq:partfunc}
\end{equation}
from which any observable will be computed. For instance, the optimal cost
(i.e. the minimum of the risk function in (\ref{eq:risk})) is computed from
\begin{equation}
  e(t,\lambda) = 
  \lim_{N\to\infty} \frac 1N
  \min_{\bf w} \text{risk}[{\bf w};N,Nt,\lambda] = 
  \lim_{N\to\infty} \frac 1N
  \lim_{\gamma\to\infty}
  \frac{-1}{\gamma} \log Z^{(N)}_\gamma[t,\lambda;\{x_i^{(\tau)}\}] \ .
  \label{eq:epsxit}
\end{equation}
It turns out that this expression depends on the actual sample (the time
series $\{x_i^{(\tau)}\}$) used to estimate the risk measure. We are mainly
interested in the average over all possible time series of this quantity,
which we assume to be narrowly distributed around its mean value. Taking the
average of eq.~(\ref{eq:epsxit}) means that we have to average the logarithm
of the partition function according to the pdf ${\sf p}
(\{x_i^{(\tau)}\})$. The so-called replica method allows to simplifiy this
task as follows. We compute $\mathbb{E}\left[Z^n\right]$ for integer $n$ and
\emph{assume} we can analytically continue this result to real $n$: then
$\mathbb{E}\left[ \log Z\right] = \lim_{n\to 0} ( \mathbb{E}
\left[Z^n\right]-1)/n $. This is the strategy that we are going to use in
the next sections and that will allow to compute the optimal portfolio.


\section{Replica analysis: Absolute Deviation \label{sect:AD}}

\begin{figure}
  \psfrag{N/T}[][][2.5]{$\displaystyle \frac 1 t$}
  \psfrag{1suradice}[][][2.5]{$q_K$ (var)}
  \psfrag{somma}[][][2.5]{$q_K $}
  \begin{center}
    \includegraphics[angle=-90,width=.49\textwidth]{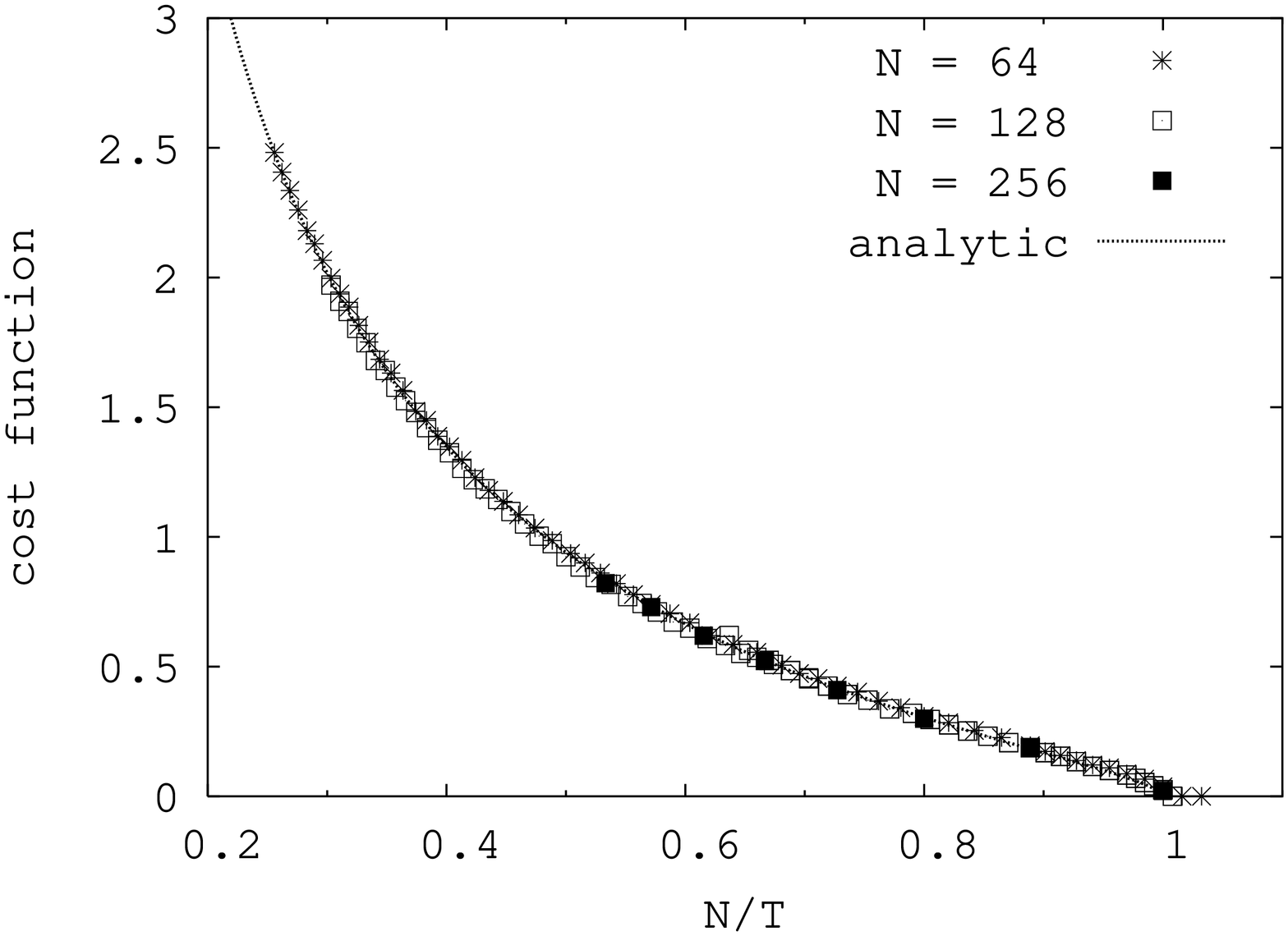}
    \includegraphics[angle=-90,width=.49\textwidth]{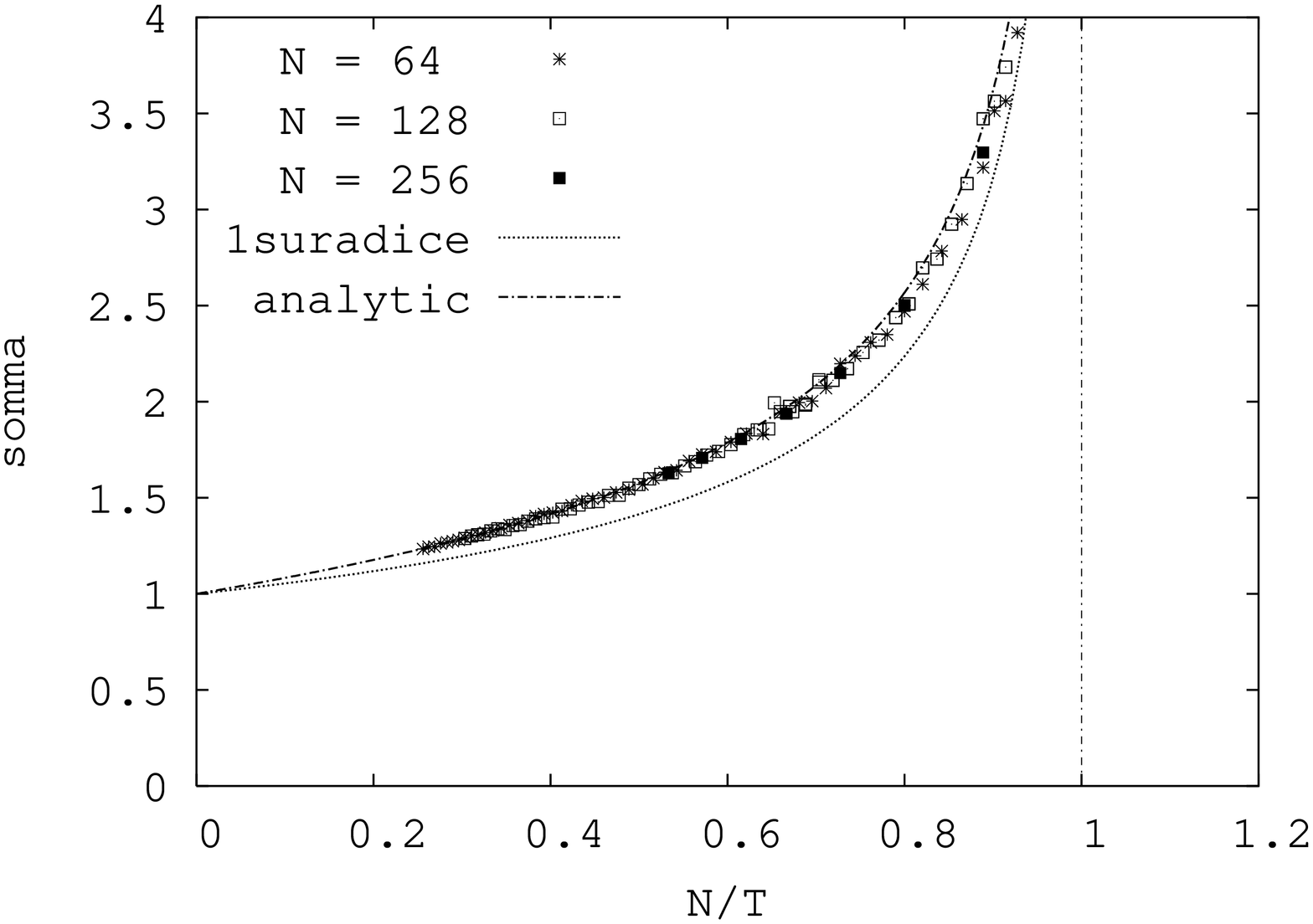}
  \end{center}
  \caption{{\bf Left:} The analytic solution $e(t)$ is compared with the
  results of numerical simulations, where the constrained optimization is
  computed directly via linear programming methods~\cite{numrec}. {\bf
  Right:} Numerical results for $\sqrt{\sum_{i=1}^N (w_i^*)^2} $ compared to
  the analytic behaviour $\sqrt{q_0'}\Delta$. The curve denoted by $q_K$
  (var) represents the behaviour of $q_K$ in the variance minimization
  problem.}
\label{fig:eneAD}
\end{figure}

The absolute deviation measure AD$\big[{\bf w}; N, T\big]$ is obtained
by choosing $\mathcal{F}_\lambda(z)=|z|$ in (\ref{eq:risk}). No other
external parameters $\lambda$ are present here. 
We assume a factorized distribution
\begin{equation}
  {\sf p}\big[\{x_i^{(\tau)}\}\big] 
  \sim \prod_{i,\tau} 
  \exp\left(-\frac {N (x_i^{(\tau)})^2}{2\sigma^2_\tau}\right) \ ,
  \label{eq:pdffact}
\end{equation}
where the volatilities $\{\sigma_\tau\}$ are distributed according to a pdf
which we do not specify for the moment. Following the replica method, we
introduce $n$ identical replicas of our portfolio and compute the average of
$Z^n$:
\begin{eqnarray}
  \mathbb{E}\left[ Z_\gamma^n(t) \right]\!
  & \sim & \!
  \int \prod_{a,b=1}^n dQ^{ab} d\hat Q^{ab}
  e^{
  N\sum_{a,b=1}^n (Q^{ab}-1) \hat Q^{ab} - \frac N2 \textrm{Tr} \log \hat Q
  -\frac T2 \textrm{Tr} \log Q 
  + \sum_\tau\log A_\gamma(\{Q^{ab}\};\sigma_\tau)
  }
  \ , \nonumber \\
  [-3mm] \label{eq:zn} \\
  &&  A_\gamma(\{Q^{ab}\};\sigma_\tau) = 
  \int \prod_{a=1}^n d u_\tau^a \exp\bigg\{
  - \frac {1}{2 \sigma^2_\tau} \sum_{ab} (Q^{-1})^{ab} u^a_\tau u^b_\tau
  -\gamma\sum_a |u_\tau^a| \bigg\}
   \nonumber \ ,
\end{eqnarray}
where we have introduced the overlap matrix 
\begin{equation}
  Q^{ab} = \frac 1N \sum_{i=1}^N w_i^a w_i^b \ , 
  \quad a,b=1,\ldots n \ ,
  \label{eq:qab}
\end{equation}
as well as its conjugate $\hat Q^{ab}$, the Lagrange multipliers introduced
to enforce (\ref{eq:qab}).  In the limit $N,T\to\infty$, $N/T=1/t$ finite,
the integral in (\ref{eq:zn}) can be solved by a saddle point method. Due to
the symmetry of the integrand by permutation of replica indices, there
exists a replica-symmetric saddle point~\cite{mepavi}: $Q^{aa}=q_1$,
$Q^{ab}=q_0$ for $a\neq b$, and the same for $\hat Q^{ab}$. We expect the
saddle point to be correct in view of the fact that the problem is
linear. Under this hypothesis, which will be only justified \emph{a
posteriori} by a direct comparison to numerical data, the replicated
partition function in (\ref{eq:zn}) gets simplified into
\begin{eqnarray}
  \mathbb{E}\left[ Z_\gamma^n (t) \right]
  & \sim & 
  \int dq_0\int  d\Delta q 
  \exp \big[Nn\ S_\gamma (q_0,\Delta q)(1+\mathcal{O}(n)) \big] \ , 
  \label{eq:SAD}\\
  S_\gamma (q_0,\Delta q) & = &
  \frac{(1-t) q_0 -1}{2 \Delta q} + \frac {1-t}{2} \log\Delta q
  + t \ \frac 1T \sum_\tau \frac 1n \log A_\gamma(q_0,\Delta q;\sigma_\tau) 
  \ , \nonumber \\
  A_\gamma(q_0,\Delta q; \sigma_\tau) & = & 
  \int  \frac{ds}{\sqrt{2\pi q_0}} e^{-s^2/2 q_0} \left[ 1 + n \int
  du\ e^{-\frac{u^2}{ 2\Delta q\sigma_\tau^2} + \frac {s\ u}{\Delta q
  \sigma_\tau} -\gamma |u| } + \mathcal{O}(n^2) \right] \ , \nonumber 
\end{eqnarray}
where $\Delta q = q_1 -q_0$ and $n$ is the number of replicas (which will
eventually go to zero).  We now assume that in the low temperature limit the
overlap fluctuations are of order $1/\gamma$ and introduce $\Delta = \gamma
\Delta q$. One can show that if $\Delta$ stays finite at low temperatures
\begin{equation}
  \lim_{n\to 0}\lim_{\gamma\to\infty} \frac 1 {n\gamma} 
  \log A_\gamma(q_0,\Delta/\gamma;\sigma_\tau) 
  = \Delta ^2 \sigma_\tau^3 \int_1^\infty ds\ 
  e^{-s^2\sigma_\tau^2\Delta^2/2q_0} (1-s)^2\ .
\end{equation}
For the sake of clarity, we focus on the simple case $\sigma_\tau = 1\
\forall \tau$. In the $\gamma\to\infty$ limit, the saddle point equations
for (\ref{eq:SAD}) are
\begin{eqnarray}
  \frac 1t & = & \textrm{erf}\left(1/\sqrt{2q'_0}\right) \ ,\label{eq:1t}\\
  \Delta & = & \left( 2 t \left[
    \frac {1-1/t}{2} q'_0 + \sqrt{\frac{q'_0}{2\pi}} e^{-1/2q'_0} 
    - \frac{(1+q'_0)}{2}\left(1-\textrm{erf}\left(1/\sqrt{2q'_0}\right)\right)
    \right] \right)^{-1/2}\ , \label{eq:Delta}
\end{eqnarray}
where $q_0 = q'_0 \Delta^2$. The minimum cost function, i.e. the average of
eq.~(\ref{eq:epsxit}), is found to be $e(t)=1/\Delta$. Notice that
(\ref{eq:1t}) only admits a solution for $t\ge 1$. There is no solution to
the minimization problem if the ratio of assets to data points, $N/T$, is
smaller than 1. On the other hand, once this condition is fulfilled, the
equation (\ref{eq:Delta}) gives a finite $\Delta$ at any $t>1$. The
asymptotic behaviour of $e(t)$ can be worked out analytically: we introduce
$\delta \equiv 1 -1/t$ and consider the limit $\delta \ll 1$. This leads to
\begin{eqnarray}
  e(t) & \simeq & \sqrt\frac{\delta}{-2\log\delta} \left(1 -
  \frac{\log\left(-\frac 4\pi \log\delta\right) }{4 \log \delta} 
  \right) \label{eq:easym}\ .
\end{eqnarray}
The full solution and a comparison with numerics are shown in
Fig.~\ref{fig:eneAD} (left). 

We now address the issue of noise sensitivity, for which a measure was
introduced in \onlinecite{pafka02}. The idea is the following: Assume you
know the true pdf of the loss (\ref{eq:pell}) and you get some optimal ${\bf
w}^{(0)}$ by minimizing the absolute deviation of $\ell$. We want to compare
the optimal risk associated to ${\bf w}^{(0)}$ with the one obtained by
optimizing (\ref{eq:risk}), i.e. the empirical estimation of the same risk
measure. A fair comparison is then $q_K-1$, with
\begin{equation}
  q^2_K (N,T) = \frac { \textrm{AD} \big[{\bf w}^*; N, T\big]} 
  {\textrm{AD} \big[{\bf w}^{(0)}; N\big]} \ ,
\end{equation}
where the $w_i^*$ refer to the portfolio obtained by minimizing
(\ref{eq:risk}). This is the quantity which we have computed by the replica
approach. In our calculation we have assumed to deal with a factorized
Gaussian distribution of returns (extensions to more realistic cases will be
presented in the next section) and it is straightforward to prove that in
this case $ q_K = \sqrt{\sum_{i=1}^N (w_i^*)^2} $. This corresponds in our
language to $\sqrt{q_0} = \sqrt{q_0'} \Delta$, which diverges like
$(1-1/t)^{-1/2}$ as $1/t \to 1^-$. Corrections to this leading behavior (which
is instead the full shape of $q_K$ in the variance minimization problem) are
needed in order to reproduce the data (right panel of Fig.~\ref{fig:eneAD}).
The comparison with the Markowitz optimal portfolio (variance minimization)
indicates that the AD measure is actually less stable to perturbations: A
geometric interpretation of this result can be found in
ref.~\onlinecite{kondor05}. Beside this fact, the interesting result is then
the existence of a well defined threshold value $t=1$ at which the
estimation error becomes infinite. This is due to the divergence of the
variance of the optimal portfolio in the regime $t<1$, where any
minimization attempt is thus totally meaningless.


\section{Expected shortfall \label{sect:ES}}

\subsection{The minimization problem}

For a fixed value of $\beta<1$ ($\beta \gtrsim 0.9$ in the interesting
cases) the expected-shortfall (ES) of a portfolio ${\bf w}$ is
obtained by choosing $\mathcal{F}(z) \propto z\theta(z-$VaR$)$ in
(\ref{eq:risk}), where VaR stands here for the
Value-at-Risk~\cite{gloriamundi}. In practice, it is computed from the
minimization of a properly chosen objective
function~\cite{rockafellar00}:
\begin{equation}
  \textrm{ES} \big[{\bf w}; N, T, \beta\big] 
  = 
  \min_v \left\{
  v + \frac 1{(1-\beta) T} \sum_{\tau=1}^T 
  \left[ 
    -v-\sum_{i=1}^N w_i x_i^{(\tau)}
    \right]^+ 
  \right\}
  \ ,
\end{equation}
where $[a]^+ \equiv (a+|a|)/2$. Optimizing the ES risk measure over all the
possible portfolios satisfying the budget constraint is equivalent to the
following linear programming problem:
\begin{itemize}
\item
  Cost function: $ E = (1-\beta)Tv+\sum_{\tau=1}^T u_\tau$\ ;
\item
  Variables: ${\bf Y} \equiv \{w_1,\ldots w_N,u_1,\ldots u_T,v\}$\ ;
\item
  Constraints: $
    u_t \ge  0 \ , \quad  
    u_t + v + \sum_{i=1}^N x_{it} w_i \ge  0 \ , \quad
    \sum_{i=1}^N w_i = N $ \ .
\end{itemize}
In a previous work~\cite{es_replicas} we solved the problem in the case
where the historical series of returns is drawn from the oversimplified
probability distribution (\ref{eq:pdffact}), with $\sigma_\tau=1\ \forall
\tau$.  Here we do a first step towards dealing with more realistic data and
assume that the series of returns can be obtained by a sequence of normal
distributions whose variances depend on time:
\begin{eqnarray}
  p\big[\{\sigma_{t}\}\big] & \sim &
  \prod_{\tau,\tau '} \exp\left(-\sigma_\tau\sigma_{\tau '} 
  G^{-1}_{\tau,\tau '}\right)
  \prod_\tau q(\sigma_\tau) \ ,
  \label{eq:probsigma}
\end{eqnarray}
for some long range correlator $G_{\tau,\tau '}$ which takes into account
volatility correlations, and $q(\sigma_\tau)$ equal e.g. to a lognormal
distribution.

\subsection{The replica solution}

A straightforward generalization of the replica calculation presented in
ref.~\onlinecite{es_replicas} (and sketched in the previous section for a
similar problem) allows to compute the average optimal cost for a given
volatility sequence $\{\sigma_1, \ldots \sigma_T\}$, in the limit when
$N,T\to\infty$ and $N/T=1/t$ stays finite. This is given by 
\begin{eqnarray}
  e(t,\beta) & = &
  \min_{v,q_0,\Delta} \left[
  \frac 1{2\Delta} + \Delta\  
  \tilde \eps (t,\beta;v,q_0|\{\sigma_\tau\}) 
  \right]
  \ , \\
  \tilde \eps (t,\beta;v,q_0|\{\sigma_\tau\}) 
  & \equiv &
  t(1-\beta)v-\frac{q_0}{2} + \frac{t}{2\sqrt{\pi}}
  \frac 1T \sum_{\tau=1}^T
  \int_{-\infty}^{+\infty} \!\!ds\  e^{-s^2} 
  g(v/\sigma_\tau+s\sqrt{2q_0};\sigma_\tau) \ ,\ \ \ \
  \label{eq:egs}
\end{eqnarray}
where $\Delta \equiv \lim_{\gamma\to\infty} \gamma\Delta q$ and the function
$g(x;\sigma)$ is equal to $x^2$ if $ -\sigma \le x < 0 $, to $-2\sigma x -
\sigma^2$ is $x < -\sigma $, and $0$ otherwise. The minimization over
$v,q_0$ implies that
\begin{equation}
  \partial \tilde \eps/\partial v =
  \partial \tilde \eps / \partial q_0 = 0  \ .
  \label{eq:speq}
\end{equation}
As discussed in \cite{es_replicas}, the problem admits a finite solution if
(\ref{eq:egs}) is minimized by a finite value of $\Delta$. The feasible
region is then defined by the condition $ \tilde \eps (t,\beta;v,q|
\{\sigma_t\}) \ge 0 $\ , where $v$ and $q_0$ satisfy (\ref{eq:speq}). This
theoretical setup suggests the following semi-analytic protocol for
determining the phase diagram of realistic portfolio optimization problems.
\begin{enumerate}
  \item
    Fix a value of $\beta \in [0,1]$, and take $N$ equal to the portfolio
    size you are interested in.
  \item
    For $T=T_\text{min}$ to $T_\text{max}$, such that $N/T \in [0.1,0.9]$,
    do the following:
    \begin{enumerate}
      \item
        Generate a sequence $\{\sigma_1,\sigma_2, \ldots \sigma_T\}$
        according to (\ref{eq:probsigma}) and compute the $\tilde \eps$
        function in (\ref{eq:egs}).
      \item
        Minimize $\tilde \eps$ with respect to $v$ and $q_0$ according to
        (\ref{eq:speq}). 
      \item
        Repeat steps (a) and (b) for $n$ samples, and compute the mean value
        $\langle \tilde \eps \rangle $.
    \end{enumerate}    
  \item
    Plot $\langle \tilde \eps \rangle$ vs. $N/T$ and find the value
    $(N/T)^*$ where this function changes its sign. 
\end{enumerate}
By repeating this procedure for several values of $\beta$ we get the phase
separation line $(N/T)^*$ vs. $\beta$.

\subsection{Results}

\begin{figure}
  \begin{center}
    \includegraphics[angle=-90,width=.7\textwidth]{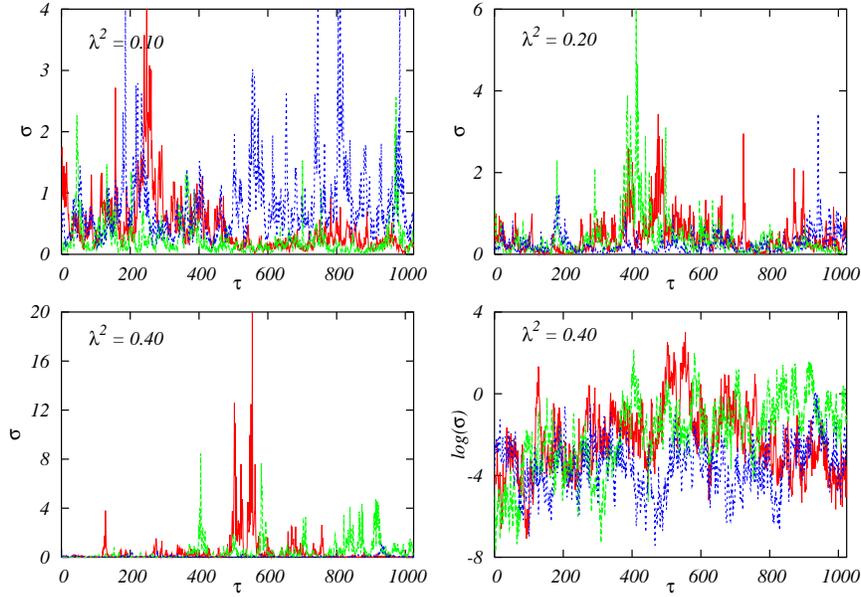}
  \end{center}
  \caption{The first three panels show 3 realizations of volatility
  sequences of length $T=1024$ according to the model
  (\ref{eq:volseq}). Different panels correspond to different values of
  $\lambda^2$. The last panel is a logarithmic representation of the
  $\lambda^2=0.40$ data. }
\label{fig:volclust}
\end{figure}

A simple way of generating realistic volatility series consists in looking
at the return time series as a cascade process~\cite{mandelbrot74}. In a
multifractal model recently introduced~\cite{muzy00} the volatility
covariance decreases logarithmically: this is achieved by letting
$\sigma_\tau = \exp \xi_\tau$, where $\xi_\tau$ are Gaussian variables and
\begin{equation}
  \langle \xi_\tau \rangle = -\lambda^2 \log T_\textrm{cut}\ , 
  \quad \langle \xi_\tau \xi_{\tau'} \rangle - \langle \xi^2_\tau \rangle
  = \lambda^2 \log \frac{T_\textrm{cut}}{1+|\tau-\tau'|} \ ,
  \label{eq:volseq}
\end{equation}
$\lambda$ quantifying volatility fluctuations (the so-called `vol of the
vol'), and $T_\textrm{cut}$ being a large cutoff. A few samples generated
according to this procedure are shown in Fig.~\ref{fig:volclust}.

\begin{figure}
  \begin{center}
    \includegraphics[angle=-90,width=.7\textwidth]{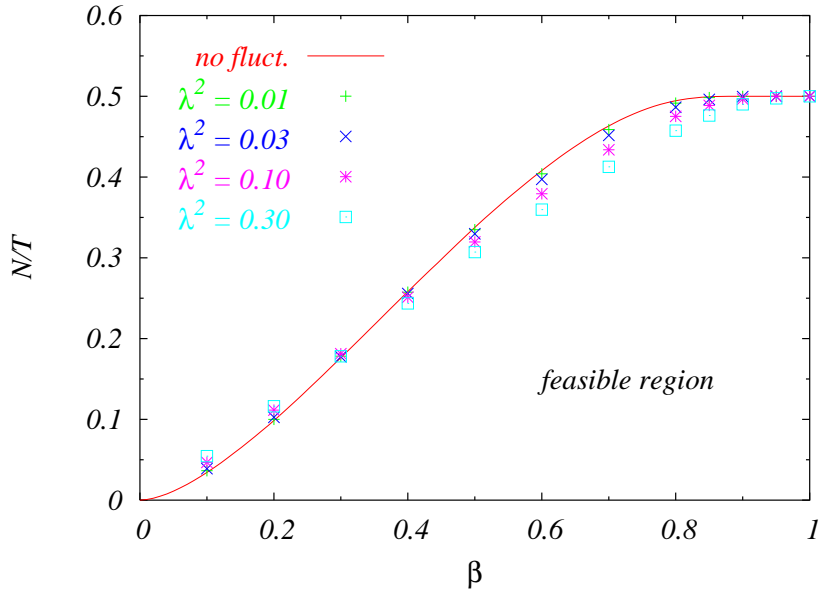}
  \end{center}
  \caption{The phase diagram corresponding to different values of the
    parameter $\lambda^2$. The full line corresponds to the absence of
    fluctuations in the volatility distributions (i.e. $\sigma_\tau=1$
    $\forall \tau$). }
\label{fig:pd}
\end{figure}

The phase diagram obtained for different values of $\lambda^2$ is shown in
Fig.~\ref{fig:pd}. A comparison with the phase diagram computed in absence
of volatility fluctuations shows that, while the precise shape of the
separating curve depend on the fine details of the volatility pdf, the main
message has not changed: There exists a regime, $N/T > (N/T)^*$, where the
small number of data with respect to the portfolio size makes the
optimization problem ill-defined. In the ``max-loss'' limit $\beta\to 1$,
where the single worst loss contributes to the risk measure, the threshold
value $(N/T)^* =0.5 $ does not seem to depend on the volatility
fluctuations. As $\beta$ gets smaller than $1$, though, the presence of
these fluctuations is such that the feasible regione becomes smaller than
the ideal multinormal case. 

\section{Conclusions \label{sect:conclu}}

In this paper we have discussed the replica approach to portfolio
optimization. The rather general formulation of the problem allows to deal
with several risk measures. We have shown here the examples of absolute
deviation, expected shortfall and max-loss (which is simply taken as the
limit case of ES). In all cases we find that the optimization problem, when
the risk measure is estimated by using time series, does not admit a
feasible solution if the ratio of assets to data points is larger than a
threshold value. As discussed in ref.~\onlinecite{kondor05}, this is a
common feature of various risk measures: the estimation error on the optimal
portfolio, originating from in-sample evaluations, diverges as a critical
value is approached. In the expected shortfall case, we have also discussed
a semi-analytic approach which is suitable for describing realistic time
series. Our results suggest that, as far as volatility clustering is taken
into account, the phase transition is still there, the only effect being the
reduction of the feasible region. As a general remark, we have shown that
the replica method may prove extremely useful in dealing with optimization
problems in risk management.

\emph{Acknowledgments.} We thank I.~Kondor and J.-P.~Bouchaud for
interesting discussions. S.~C. is supported by EC through the network MTR
2002-00319, STIPCO.

\end{document}